\documentclass[reprint,amsmath,amssymb,prl]{revtex4-2}
\usepackage{graphicx,subfigure,bm,amssymb,amsmath,hyperref,dcolumn}
\usepackage{color,multirow,supertabular,float}
\usepackage{mathrsfs}

\usepackage{physics}
\usepackage{amsmath}
\usepackage{tikz}
\usepackage{mathdots}
\usepackage{yhmath}
\usepackage{cancel}
\usepackage{color}
\usepackage{siunitx}
\usepackage{array}
\usepackage{multirow}
\usepackage{amssymb}
\usepackage{gensymb}
\usepackage{tabularx}
\usepackage{extarrows}
\usepackage{booktabs}
\usetikzlibrary{fadings}
\usetikzlibrary{patterns}
\usetikzlibrary{shadows.blur}
\usetikzlibrary{shapes}

\begin{abstract}
We present high-precision \emph{ab initio} calculations of the four-point vertex function for the three-dimensional uniform electron gas using variational diagrammatic Monte Carlo. From these results, we extract Landau parameters that reveal a density-driven crossover from underscreening to overscreening, and obtain the full two-electron scattering amplitude on the Fermi surface with controlled accuracy. A residual analysis of the scattering amplitude against the charge-channel Kukkonen--Overhauser (KO$^+$) interaction shows that only a minimal s-wave correction in the antiparallel-spin channel is needed, defining the sKO$^+$ ansatz: KO$^+$ within the local-density approximation plus this short-range correction. Using both our direct VDMC amplitudes and the sKO$^+$ ansatz, we compute the electron-electron contribution to the thermal resistivity, obtaining quantitative agreement with experiments on simple metals (Al, Na, K, Rb). sKO$^+$ thus provides a controlled UEG-based effective interaction for simple-metal transport and future first-principles extensions.
\end{abstract}

\begin{document}
\title{Two-Electron Correlations in the Metallic Electron Gas}
\author{Zhiyi Li$^{1,2}$}
	\thanks{These two authors contributed equally to this paper.}
 \author{Pengcheng Hou$^{2}$}
 	\thanks{These two authors contributed equally to this paper.}
 \author{Bao-Zong Wang$^{3}$}
 \email{bzwang.phys@gmail.com}
	\author{Youjin Deng$^{1,2,4}$}
        \email{yjdeng@ustc.edu.cn}
	\author{Kun Chen$^{5}$}
        \email{chenkun@itp.ac.cn}
	\affiliation{$^{1}$ Department of Modern Physics, University of Science and Technology of China, Hefei, Anhui 230026, China}
	\affiliation{$^{2}$ Hefei National Laboratory, University of Science and Technology of China, Hefei 230088, China}
    \affiliation{$^{3}$ International Center for Quantum Materials, School of Physics, Peking University, Beijing 100871, China}
        \affiliation{$^{4}$ Hefei National Research Center for Physical Sciences at the Microscale and School of Physical Sciences, University of Science and Technology of China, Hefei 230026, China}
        \affiliation{$^{5}$CAS Key Laboratory of Theoretical Physics, Institute of Theoretical Physics, Chinese Academy of Sciences, Beijing 100190, China}
\maketitle

\emph{Introduction.---}
The collective behavior of interacting electrons gives rise to some of the most profound and challenging problems in condensed matter physics~\cite{alexandradinata2024futurecorrelatedelectronproblem}, including high-temperature superconductivity~\cite{HTCSC}, strange metallicity~\cite{greene2020strange}, and the emergence of topological phases~\cite{Toplogphase}. The uniform electron gas (UEG)---an idealized model of electrons moving in a uniform neutralizing background---provides the quintessential canvas for understanding correlations originating solely from the Coulomb interaction~\cite{UEGintro}. While the single-particle properties of this system have been extensively studied~\cite{mahan2013many}, a comprehensive understanding of two-electron correlations has remained elusive despite their fundamental importance.

Two-electron correlations on the Fermi surface are encoded in the four-point vertex function, which simultaneously controls Fermi-liquid parameters~\cite{abrikosov2012methods,negele2018quantum}, pairing instabilities~\cite{PhysRev.108.1175}, and the electron-electron scattering rates that govern transport phenomena~\cite{PhysRevLett.21.279}.
A first-principles determination has historically been impeded by its complex dependence on multiple momentum and frequency variables. The immense computational cost associated with this high dimensionality poses a severe challenge, forcing many theoretical frameworks to rely on approximations, such as low-order truncations in diagrammatic methods or the local approximations in dynamical mean-field extensions and post-GW schemes~\cite{dMFT1,DMFT2,cunningham2025manybodyvertexeffectstimedependent,Zang_2024}. Conventional quantum Monte Carlo (QMC) techniques face a complementary limitation. They accurately access ground-state energies and selected correlations, but not the vertex function directly. Extracting it requires four-point correlations that are difficult to resolve with sufficient accuracy in standard QMC.
This theoretical gap is particularly pressing today, as advances in two-electron photoemission and multidimensional electronic spectroscopies begin to access many-electron correlations experimentally~\cite{PPS,Su_proposeARPES,fresch2023two}.

\begin{figure}
    \centering
    \includegraphics[width=0.8\linewidth]{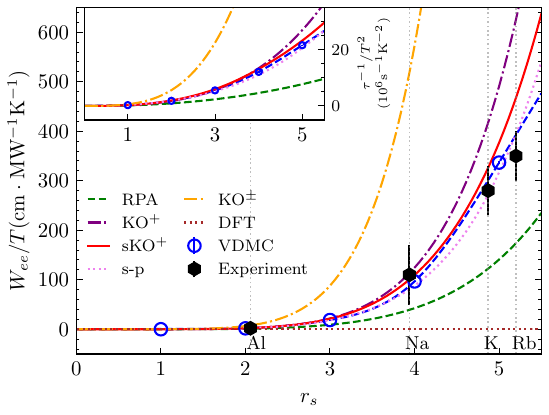}
    \caption{The electron-electron contribution to the thermal resistivity $W_{ee}$ as a function of density parameter $r_s$ in the metallic regime, computed from our variational diagrammatic Monte Carlo (VDMC) results (blue circles) and from several theoretical models, including the random phase approximation (RPA), the s-p approximation~\cite{transportcoeff1969,MacDonald_1980}, the original Kukkonen--Overhauser interaction~\cite{PhysRevB.20.550} (KO$^\pm$), and the charge-channel-only form (KO$^+$). For comparison, we plot experimental data (black circles) for simple metals~\cite{Cook1972,Cook1973,Cook1979,Uher2004}. The solid red line shows our proposed sKO$^+$ model, namely KO$^+$ supplemented by s-wave corrections, which agrees well with the VDMC results and experimental measurements. The inset shows the electron-electron scattering rate.}
    \label{fig:Transport}
\end{figure}

The consequences of this theoretical gap are particularly stark in the study of transport phenomena, presenting a long-standing challenge to the first-principles description of thermal conductivity in simple metals. Here, electron-electron scattering is understood to be a primary contributor to the experimentally observed deviations from the Wiedemann--Franz law~\cite{Kresistivity1979,Kresistivity1980,Cook1972,Cook1973,Cook1979,Uher2004}, an effect that conventional density functional theory (DFT) struggles to account for~\cite{PhysRevE.101.053204}. Previous attempts to resolve this issue using many-body theories, such as the random phase approximation (RPA)~\cite{lindhard1954properties,KukkonenRPA} and the Kukkonen--Overhauser (KO) interaction~\cite{KOscattering,PhysRevB.20.550}, have also resulted in thermal resistivity values that disagree with experimental measurements. Consequently, finding a theoretical ansatz for the effective electron-electron interaction that yields precise quantitative agreement with experiment remains an open challenge and makes this problem directly relevant to transport theory and experiments in simple metals.

In this work, we address this challenge using the variational diagrammatic Monte Carlo (VDMC) method~\cite{chen2019combined,haule2022single,diagMC1, diagMC2}, which combines a variational principle with high-order, controlled stochastic summation of Feynman diagrams~\cite{kozik, diagMC4, boldDiagMC, rossi2018, van2012feynman,diagMC-FiniteSystem,wang2025variationaldiagrammaticmontecarlobuilt,hou2024feynman,spinXC,WDM_XC,chen2024partial}. For the 3D UEG, we compute the four-point vertex up to sixth order, comprehensively characterizing two-electron correlations and extracting the Landau parameters $F_{0,1,2}^{s,a}$. This provides a long-sought, first-principles quantitative determination of these key Fermi-liquid parameters. Our results reveal a crossover from underscreening to overscreening~\cite{Takada_rsc3d,Matsudaexp2007,PhysRevB.56.4872}, where collective screening drives the static dielectric function negative, effectively inverting the test-charge interaction from repulsive to attractive. This transition is characterized by the Landau parameter $F_0^s$ approaching $-1$, signaling a divergence in electronic compressibility and the onset of a strongly correlated regime. 

Furthermore, we calculate the full electron-electron scattering amplitude with high precision across the entire Fermi surface. Guided by these numerical results, we introduce the following effective interaction:
\begin{equation}
\begin{aligned}
    R^{\sigma\sigma'}_{\mathbf{k}_1\mathbf{k}_2\mathbf{q}}\!
    &=\! \frac{v_q+f_{\rm XC}^+}{1-[v_q+f_{\rm XC}^+]\Pi_0(q)}
    \!-\! f_{\rm XC}^+\!+\!\delta R^{\sigma\sigma'}, \\
    \delta R^{\sigma\sigma'}\!\! & =\! (\!1\!-\!\delta_{\sigma\sigma'}\!)\!\left[C_0 \!+\! C_2\frac{(\mathbf{k}\!_1\!-\!\mathbf{k}\!_2)^2\!+\!(\mathbf{k}\!_1\!-\!\mathbf{k}\!_2\!-\!2\mathbf{q})^2}{2}\right],
\end{aligned}
\label{effective interaction}
\end{equation}
Here $f_{\rm XC}^+$ denotes the symmetric exchange-correlation (XC) kernel. The XC kernels discussed in this work are treated within LDA by replacing $f_{\rm XC}^{\pm}(q)$ with their static, long-wavelength values $f_{\rm XC}^{\pm}(0)$. Specifically, this ansatz is constructed by starting from the charge-channel KO interaction and adding a small s-wave correction $\delta R$ acting only in the antiparallel-spin channel, parametrized by $C_0$ and $C_2$ and associated with the s-wave scattering length and effective range. By setting $\delta R=0$, Eq.~\eqref{effective interaction} recovers the charge-channel KO model, KO$^+$. This ansatz resolves the discrepancies found in widely used approximations such as RPA and KO. It provides a quantitatively accurate effective interaction on the Fermi surface, extending naturally to finite momentum through the closed-form structure inherited from KO$^+$.

Crucially, our first-principles results for the electron-electron scattering amplitude enable a direct calculation of the thermal resistivity, resolving the long-standing discrepancy between previous theoretical approximations and experimental measurements. As shown in Fig.~\ref{fig:Transport}, the resistivity derived from our calculated scattering amplitude is in good agreement with experimental data~\cite{Cook1972,Cook1973,Cook1979,Uher2004}. Moreover, the thermal resistivity calculated using our proposed sKO$^+$ model also closely matches the experimental values, confirming the validity of our comprehensive approach. Beyond this validation, the closed-form sKO$^+$ provides a UEG-based input for downstream first-principles methods, with concrete applications to time-dependent DFT (TDDFT) kernels, Eliashberg- and Boltzmann-type transport calculations, and effective-field-theory constructions discussed below.

\emph{Methods.---}
We employ the VDMC framework~\cite{chen2019combined,haule2022single,diagMC1, diagMC2, kozik, diagMC4, boldDiagMC, rossi2018, van2012feynman,diagMC-FiniteSystem,wang2025variationaldiagrammaticmontecarlobuilt}, which combines a robust variational principle with stochastic sampling of Feynman diagrams. This approach enables controlled evaluation of high-order perturbative contributions to the four-point vertex function of the UEG.

The microscopic model is the standard UEG Hamiltonian
\begin{align}
    H =& \sum_{\mathbf{k}\sigma}(\mathbf{k}^2-\mu)\psi^\dagger_{\mathbf{k}\sigma}\psi_{\mathbf{k}\sigma} \notag \\
    &+\frac{1}{2}\sum_{\substack{\mathbf{q}\neq0\\ \mathbf{k}\mathbf{k}'\sigma\sigma'}}
    v_q\psi^\dagger_{\mathbf{k}+\mathbf{q}\sigma}\psi^\dagger_{\mathbf{k}'-\mathbf{q}\sigma'}
    \psi_{\mathbf{k}'\sigma'}\psi_{\mathbf{k}\sigma},
\end{align}
where $\psi_{\mathbf{k}\sigma}$ and $\psi^\dagger_{\mathbf{k}\sigma}$ are the annihilation and creation operators for an electron with momentum $\mathbf{k}$ and spin $\sigma$, $\mu$ is the chemical potential, and $v_q = 4\pi e^2/q^2$ is the Coulomb interaction.

The central idea is to reorganize many-body perturbation theory around a variationally optimized screened interaction. Counterterms are introduced to preserve the exact physics of the original Hamiltonian, ensuring that the expansion remains unbiased. The diagrammatic series is then sampled as a computational graph, where counterterm contributions are obtained through Taylor-mode automatic differentiation. Further technical details of the VDMC method are given in Sec.~SI of the Supplemental Material.

Our implementation incorporates automated high-order renormalization and GPU-accelerated integration routines~\cite{hou2024feynman}, allowing efficient evaluation of high-dimensional diagrams up to sixth order. This combination provides unprecedented computational reach for effective field theories in strongly interacting systems.

Systematic cancellation between same-order expansion terms generated when the constituent quantities of a physical observable are multiplied accelerates the convergence of the final result. For the Landau parameters, this mechanism is shown explicitly in Eqs.~S2--S3 and Fig.~S1 of the Supplemental Material: the combined observable converges faster than the separate quasiparticle-weight, effective-mass, and vertex ingredients. For instance, at $r_s = 4$ the combined parameters converge by sixth order, with uncertainties estimated from the spread of the last three expansion orders.

\begin{figure}
    \centering
    \includegraphics[width=0.98\linewidth]{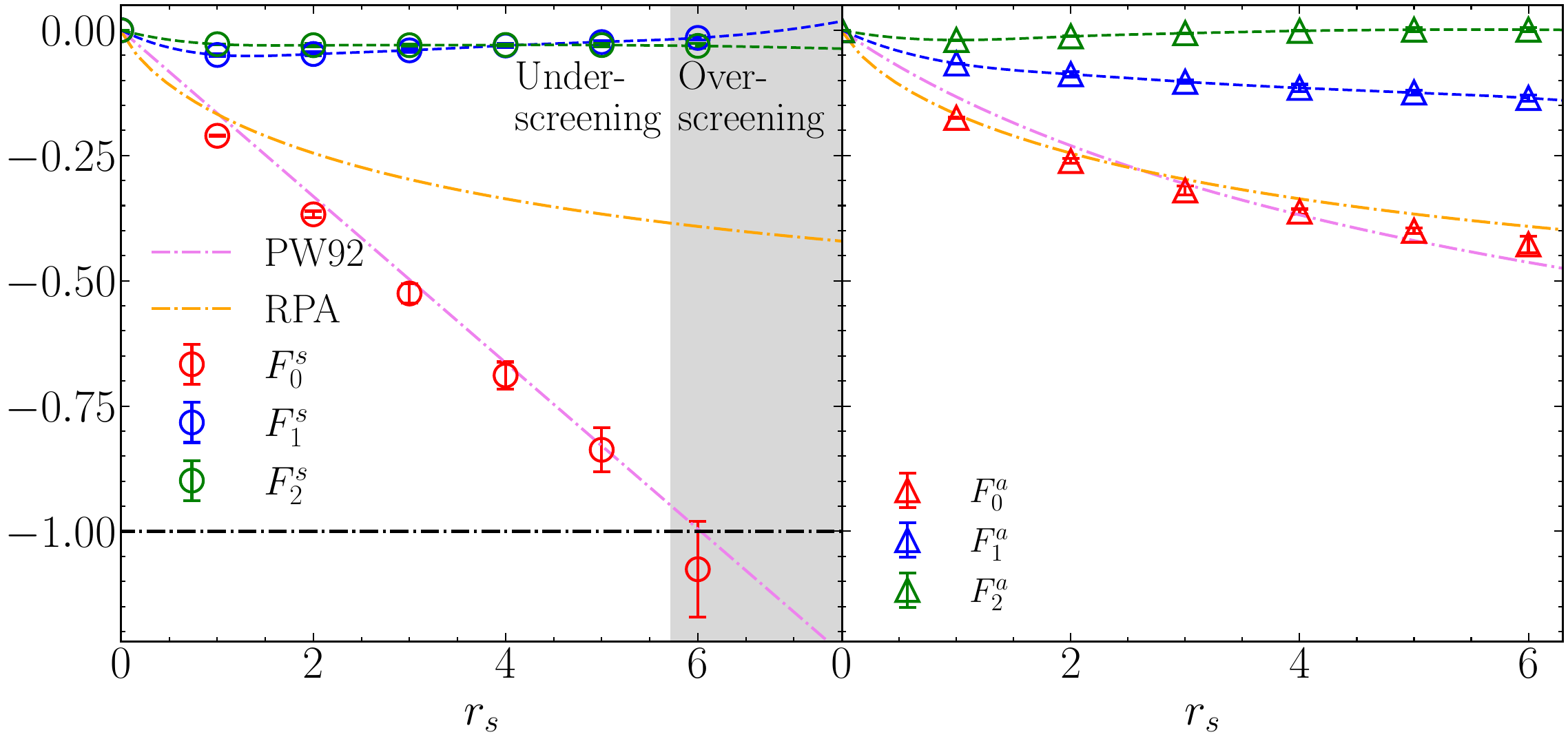}
    \caption{Numerical results of the symmetric (left panel) and antisymmetric (right panel) Landau parameters versus $r_s$, confirming the rapid convergence of the Legendre expansion. The linear decrease of $F_0^s$ with $r_s$, which approaches $-1$ at $r_s^c \approx 5.7$, reveals the crossover of the Coulomb screening effect from underscreening to overscreening~\cite{Takada_rsc3d}. For comparison, the orange dash-dot line shows the $F_0$ calculated from the RPA vertex. The violet dash-dot line presents the dimensionless static XC kernel $f_{\rm XC}(q=0)N_F$ for the UEG derived from the PW92 energy functional~\cite{PW92}, demonstrating the approximate equality of the XC kernel and the zeroth Landau parameter. The corresponding numerical data are listed in Table S1 of the Supplemental Material.}
    \label{fig:LPvsrs}
\end{figure}
\emph{Landau parameters and screening crossover.---}
The 3D UEG is characterized by the density parameter, the Wigner--Seitz radius $r_s= (\frac{3}{4\pi n})^{1/3}$ and the reduced temperature $\Theta = T/T_F$. We investigate $r_s=1$--6 in the low-temperature, degenerate Fermi-liquid regime, using $\Theta = 0.025$ and $0.05$. The corresponding physical temperature scale is set by the Fermi temperature $T_F=\hbar^2 k_F^2/(2m_ek_B)$, where the Fermi momentum is $k_F=(9\pi/4)^{1/3}/(a_0r_s)$, giving $T_F\simeq 5.82\times10^5/r_s^2$ K. This scale shows that the lowest temperatures reached in our calculations are of order room temperature even at $r_s = 6$.
Within Landau-Silin theory~\cite{silin1958theory,silin1958theory2}, the effective short-range interaction between quasiparticles on the Fermi surface $f_{\sigma\sigma'}(\theta)$, which encapsulates many-body correlations, is obtained from the small-frequency and long-wavelength limit of the four-point vertex diagram $\Gamma_4$ as
\begin{equation}
    f_{\sigma\sigma^\prime}(\theta)=z^2\lim_{\mathbf{q}\to 0} [\Gamma_4^{\sigma\sigma'}(\mathbf{k_1}\omega_{-1},\mathbf{k}_2\omega_0;\mathbf{k_1+q}\omega_{0})-v_q].
\end{equation}
Here, $\theta$ is the angle between the incoming momenta $\mathbf{k}_1$ and $\mathbf{k}_2$ on the Fermi surface, $\omega_0 = i\pi T$ and $\omega_{-1} = -i\pi T$ are fermionic Matsubara frequencies, and $z$ is the quasiparticle renormalization factor derived from the self-energy~\cite{hou2024feynman}. Meanwhile, the two-electron scattering amplitude, $a_{\sigma\sigma'}(\theta,\phi)$, is related to the four-point vertex for elastic scattering of quasiparticles on the Fermi shell as:
\begin{equation}
    a_{\sigma\sigma'}(\theta,\phi) = z^2 \Gamma_4^{\sigma\sigma'}(\mathbf{k}_1\omega_0 , \mathbf{k}_2\omega_0; \mathbf{k}_3\omega_0),
    \label{eq:a_definition}
\end{equation}
where $\phi$ is the angle between the plane defined by $(\mathbf{k}_1, \mathbf{k}_2)$ and the plane defined by the outgoing momenta $(\mathbf{k_3}, \mathbf{k_4})$.

Figure~\ref{fig:LPvsrs} shows the dimensionless spin-symmetric and antisymmetric Landau parameters $F_l^{s,a}$ ($l=0,1,2$) as functions of $r_s$. These parameters, which characterize fundamental many-body interactions in the electron gas, are defined as Legendre components of the dimensionless quasiparticle interaction $F^{s,a}(\theta) = (F_{\uparrow\uparrow}(\theta)\pm F_{\uparrow\downarrow}(\theta))/2$ with $F_{\sigma\sigma'}(\theta) = N_F^*f_{\sigma\sigma'}(\theta)$, where $N_F^* = \frac{m^*}{m}N_F$ is the renormalized density of states at the Fermi surface.

One pivotal observation is the linear decrease of the compressibility-related parameter: $F_0^s \approx -0.17 r_s$. This behavior drives the Coulomb screening effect from ``underscreening'' to ``overscreening'', with the crossover occurring at $F_0^s = -1$ where the inverse compressibility (equivalently, the bulk modulus) vanishes and the ion-ion interaction becomes effectively attractive at large distances~\cite{maebashi2007pseudoquantumcriticalityelectronliquids,Matsudaexp2007}. Our calculations show that this crossover occurs near $r_s^c \approx 5.7$, close to the previous theoretical prediction of $r_s = 5.25$ in Ref.~\cite{Takada_rsc3d}~\footnote{A further discussion of this crossover phenomenon, including numerical results for the density response function and the ion-ion interaction, is provided in Sec.~IV of the Supplemental Material.}.
Importantly, the higher-order Landau parameters remain significantly smaller than $F_0^s$, demonstrating the rapid convergence of the Legendre expansion.  For comparison, Fig.~\ref{fig:LPvsrs} also shows $F_0^s$ obtained from the RPA vertex and the dimensionless zero-momentum static XC kernel $f_{\text{XC}}^{\pm}(q=0)N_F$ derived from the PW92 energy functional~\cite{PW92}.
The PW92 XC kernels agree well with our numerically extracted $F_0^{s,a}$, confirming the approximate equality between the static zero-transfer-momentum XC kernel and the corresponding Landau parameters as
\begin{equation}
    f_{\rm XC}^{\pm}(q=0)N_F \approx F_0^{s,a}  ,~\label{FXCandF}
\end{equation}
for $r_s$ in the metallic regime---a relation specific to the homogeneous electron liquid and not generally shared by arbitrary Fermi liquids. The theoretical insights of this relation will be discussed in Sec.~III in the Supplemental Material.

In addition, the parameter $F_1^s$ plays a crucial role in determining the effective mass $m^*$ through the relation $m^*/m = 1+F_1^s$, where $m$ represents the bare electron mass. Our finding that $F_1^s$ remains small ($|F_1^s|\lesssim 0.05$) throughout the metallic density regime confirms the variational Monte Carlo and VDMC results~\cite{TakadaEM,markus2023,haule2022single} that the effective mass approximately equals the bare mass in this regime.

\begin{figure}
    \centering
    \includegraphics[width=0.8\linewidth]{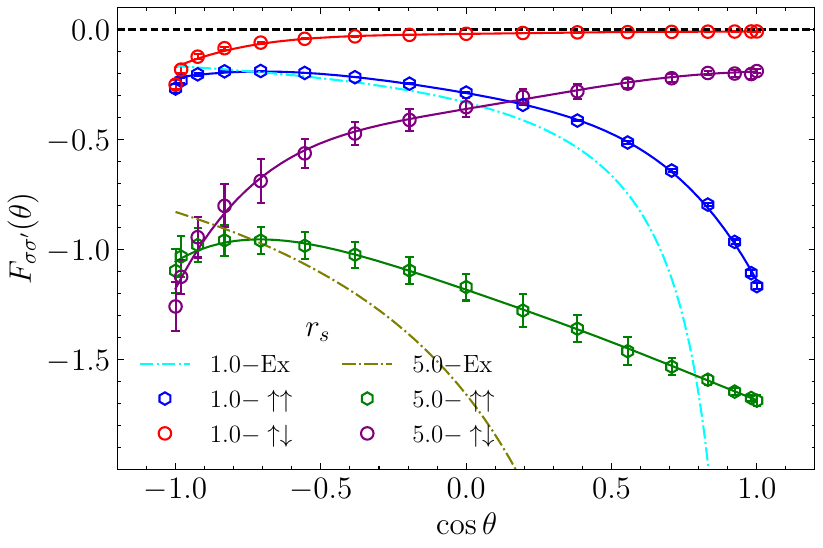}
    \caption{Angle-resolved Landau quasiparticle interaction for $r_s =1,5$ with parallel and antiparallel spins. The dotted-dashed line represents the bare exchange Coulomb interaction. The larger deviation from the bare interaction for $r_s=5$ indicates that Coulomb screening increases as the electron density decreases.}
    \label{fig:LQI-AR}
\end{figure}

\begin{figure*}
    \centering
    \includegraphics[width=0.7\linewidth]{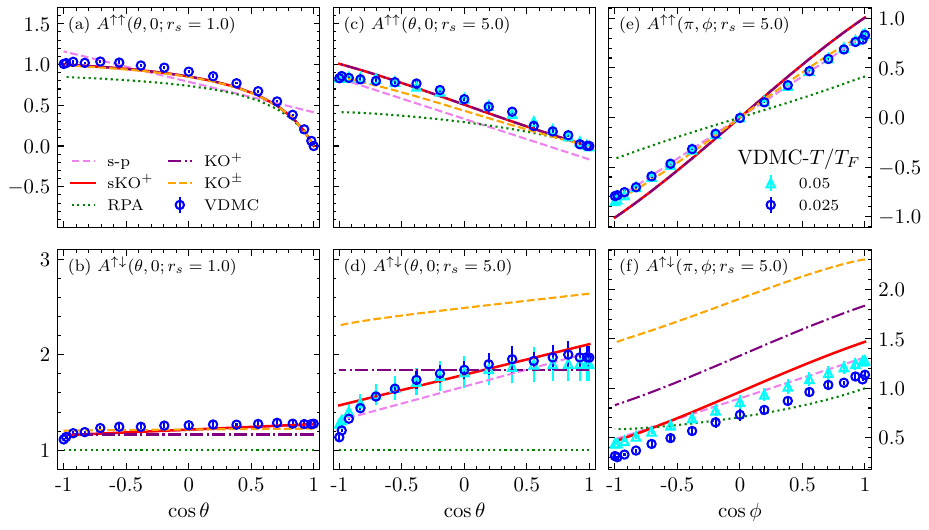}
    \caption{The dimensionless two-electron scattering amplitude $A^{\sigma\sigma'}(\theta, \phi)$ for parallel ($\uparrow\uparrow$, top row) and antiparallel ($\uparrow\downarrow$, bottom row) spins at $r_s = 1$ and $r_s=5$ at $T/T_F = 0.025$. Panels (a-d) show the forward scattering channel ($\phi=0$). Panels (e-f) show the Cooper channel ($\theta=\pi$). Our VDMC results (symbols) are compared against several theoretical models. The solid curves show the sKO$^+$ ansatz obtained from the residual analysis of our VDMC data relative to the KO$^+$ interaction (see Fig.~\ref{fig:residual_expansion} and main text), agreeing well with the numerical results. For $r_s = 5$, numerical results at $T/T_F = 0.05$ are also plotted. The divergence of the scattering amplitude at both low temperatures in the Cooper channel is consistent with the expected logarithmic divergence of the four-point vertex function as $T\to 0$~\cite{PhysRevB.106.L220502}.}
    \label{fig:scattering amplitude}
\end{figure*}

Figure~\ref{fig:LQI-AR} displays the angle-resolved Landau quasiparticle interaction $F_{\sigma\sigma'}(\theta)$ versus $\cos\theta$ at $r_s=1$ and $5$ for both parallel ($\uparrow\uparrow$) and antiparallel spin ($\uparrow\downarrow$) configurations, together with the bare exchange Coulomb interaction $W_{\rm Ex}(\mathbf{k}_1,\mathbf{k}_2) = -N_Fv(|\mathbf{k}_1-\mathbf{k}_2|)$ for reference. Two features of the many body effect are observed through our data. First, in the parallel-spin channel, $F_{\uparrow\uparrow}(\theta)$ deviates from the bare exchange primarily at small angles at $r_s=1$ but becomes smoothly angle-dependent at $r_s=5$ with deviations across all angles, indicating enhanced screening and an increasingly local quasiparticle interaction as the density decreases---a trend that naturally justifies an LDA treatment of the charge channel. Second, the antiparallel-spin channel $F_{\uparrow\downarrow}(\theta)$, which receives no bare short-range contribution, develops a pronounced minimum at $\theta=\pi$ associated with dynamical screening that drives Cooper-pairing instability~\cite{PhysRevB.106.L220502}.

\emph{Scattering amplitude and the sKO$^+$ ansatz.---}
Figure~\ref{fig:scattering amplitude} presents the dimensionless two-electron scattering amplitude, $A^{\sigma\sigma'}(\theta, \phi) = N_F^*a_{\sigma\sigma'}(\theta,\phi)$, for the UEG. We analyze the forward-scattering ($\phi=0$) and Cooper ($\theta=\pi$) channels at $r_s=1$ and 5 for $T/T_F = 0.025$, and include $T/T_F = 0.05$ for $r_s=5$. Comparison of these datasets demonstrates that our results have reached the zero-temperature limit within error bars, except for the Cooper channel divergence caused by the Tolmachev singularity~\cite{PhysRevB.106.L220502}. Crucially, as this singularity is confined to a phase space of measure zero, it has a negligible impact on Fermi-surface integrals in general. Based on these validated numerical results, we propose the sKO$^+$ ansatz: it is based on the charge KO interaction within LDA, which treats the symmetric (antisymmetric) XC kernel $f^\pm_{\rm XC}(q) \approx f_{\rm XC}^\pm(0)$($:=f_{\rm XC}^\pm$ as mentioned previously), supplemented by an s-wave correction $\delta R$ for the antiparallel spin configuration as presented in Eqs.~\eqref{effective interaction}.


To motivate the structure of the proposed sKO$^+$ ansatz, we first benchmark common theoretical models against our high-precision VDMC data in Fig.~\ref{fig:scattering amplitude}. (i) RPA: the simplest case of Eq.~\eqref{effective interaction}, where the XC effect is entirely neglected as $f_{\rm XC}^+(0)=0$ and $\delta R=0$, is clearly inadequate, especially in the low-density regime ($r_s \sim 5$). (ii) KO$^+$: the charge-channel KO interaction, which corresponds to $\delta R=0$ in Eq.~\eqref{effective interaction}, agrees well with numerical results in the forward-scattering channel but deviates significantly approaching the Cooper channel. (iii) KO$^\pm$: the full KO interaction, which further incorporates the spin-channel interaction $R_- = \hat{\sigma}\cdot \hat{\sigma}' \left[\frac{f_{\rm XC}^-}{1-f_{\rm XC}^-\Pi_0(q)} - f_{\rm XC}^-\right]$, performs worse as $r_s$ increases, particularly for antiparallel spins (Fig.~\ref{fig:scattering amplitude}(d,f)). The superior performance of KO$^+$ over KO$^\pm$ suggests that spin interactions are implicitly captured via the exchange component of the charge interaction, without requiring explicit spin-channel terms. (iv) The s-p approximation~\cite{transportcoeff1969,MacDonald_1980}: it truncates scattering to s- and p-wave channels, which matches our numerical data well, confirming that higher-order partial waves are negligible. However, this approximation is restricted to the Fermi surface and fails to provide a complete effective interaction at other energy scales.

The relatively good performance of KO$^+$ indicates that the dominant long-range charge screening is already captured, so the remaining discrepancy can be treated as a residual short-range contribution. We therefore analyze the residual scattering amplitude
\begin{equation}
\delta A^{\sigma\sigma'}(\theta,\phi)=A_{\rm VDMC}^{\sigma\sigma'}(\theta,\phi)-A_{\text{KO}^+}^{\sigma\sigma'}(\theta,\phi),
\end{equation}
and expand it in the orthogonal Fermi-surface basis~\cite{pfitzner1985general}
\begin{equation}
\delta A^{\sigma\sigma'}(\theta,\phi)=\sum_{k=0}^{\infty}\sum_{l=0}^{k}\delta A_{lk}^{\sigma\sigma'}X_{lk}(\theta,\phi).
\end{equation}

\begin{figure}[t]
    \centering
    \includegraphics[width=0.95\linewidth]{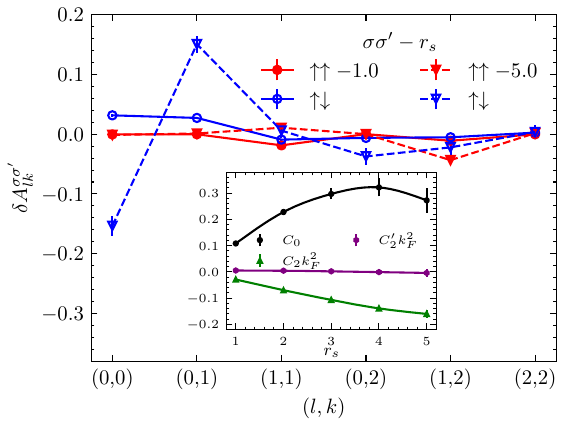}
    \caption{General polynomial expansion~\cite{pfitzner1985general} of the residual scattering amplitude beyond the KO$^+$ ansatz. The residual is negligible for parallel spins and is dominated by the s-wave components for antiparallel spins. The inset sketches the resulting parametrization of $\delta R$ in Eq.~\eqref{effective interaction}, where $C_0$ and $C_2$ correspond to the s-wave scattering length and effective range, while the higher-order p-wave correction $C_2'$ is negligible.}
    \label{fig:residual_expansion}
\end{figure}

As summarized in Fig.~\ref{fig:residual_expansion} (with further details in Sec.~V of the Supplemental Material), this expansion shows that the residual is negligible for parallel spins and is dominated by the $(l,k)=(0,0)$ and $(0,1)$ components for antiparallel spins. Guided by this clear structure, we formulate the sKO$^+$ effective interaction in Eqs.~\eqref{effective interaction}. This construction keeps the exact long-range Coulomb screening inherited from KO$^+$ while correcting the antiparallel-spin deficit through a minimal contact-interaction ansatz inspired by effective field theory for dilute Fermi gases~\cite{HAMMER2000277,WELLENHOFER2020135247} with $C_0$ and $C_2$ related to the s-wave scattering length and effective range. As shown by the solid curves in Fig.~\ref{fig:scattering amplitude}, the resulting sKO$^+$ reproduces the full VDMC scattering amplitude on the Fermi surface across all tested densities, channels, and spin configurations, thereby resolving the Fermi-surface deficiencies of the conventional models.

\emph{Thermal resistivity in simple metals.---}
We now apply our calculated scattering amplitudes to compute macroscopic transport properties~\cite{mahan2013many,PhysRev.185.323,PhysRevLett.21.279}, enabling a direct validation of our numerical results against experimental benchmarks. In simple metals, deviations from the Wiedemann--Franz law---which asserts that the electrical resistivity $\rho$ and thermal resistivity $W$ relate to a universal constant $L_0$ via the equation $\rho/(WT) = L_0$---are experimentally observed, particularly above the Debye temperature~\cite{kittel2018introduction}. This deviation manifests as an additional thermal resistance with a linear temperature dependence, identified as the electron--electron scattering contribution, $W_{ee}$~\cite{Cook1972,Cook1973,Cook1979,Uher2004}. However, a precise theoretical description has remained elusive: conventional first-principles DFT typically struggles to capture such dynamical scattering processes, while standard many-body approximations like RPA and the original KO$^\pm$ interaction fail to reproduce the experimental magnitude---yielding significant underestimates and overestimates, respectively (Fig.~\ref{fig:Transport}). By contrast, utilizing the full angular dependence of our \emph{ab initio} scattering amplitudes~\footnote{See Sec.~SVI in the Supplemental Material}, our VDMC calculations achieve excellent agreement with measurements for simple metals (Al, Na, K, Rb). For these nearly-free-electron metals, band-mass and core-polarization corrections are small. Therefore, the crystal structures and the DFT bands are not included at this stage as a density-parametrized UEG benchmark and these metals are only represented by the corresponding valence-electron density parameter $r_s$, allowing our results to be compared directly with the measured electron--electron thermal resistivity. Remarkably, this agreement is faithfully reproduced by the computationally efficient sKO$^+$ ansatz, highlighting its promise for incorporating accurate electron-electron scattering into future material-specific first-principles transport calculations. Complementing these macroscopic resistivity results, we also present the quasiparticle scattering rate in the inset of Fig.~\ref{fig:Transport}, providing the fundamental timescale governing these transport dynamics~\cite{mahan2013many,PhysRevLett.21.279}.

\emph{Discussion.---} In summary, we have presented high-precision \emph{ab initio} calculations of the four-point vertex function of the 3D UEG using our developed VDMC framework. The extracted Landau parameters validate LDA and reveal how the dominant $F_0^s$ drives the system from underscreening toward overscreening as the density decreases. Guided by a residual analysis of our scattering amplitudes against the KO$^+$ interaction, we have established the sKO$^+$ ansatz: a charge-channel KO interaction in LDA supplemented by a minimal antiparallel-spin s-wave correction. The sufficiency of such short-range corrections parallels recent findings that imperfect screening at short distances is essential for capturing the physics of the low-density regime~\cite{PhysRevLett.134.046402}.

The excellent agreement among experimental thermal-resistivity measurements, our direct VDMC calculations, and the sKO$^+$ ansatz demonstrates that this simple effective interaction is both quantitatively accurate and predictive. Its closed form is built from the static XC kernel and two short-range coefficients tabulated from UEG data, making it straightforward to use as an electron--electron kernel in TDDFT or related dielectric-response calculations, where the long-wavelength charge response must be retained while local correlation effects are improved. Because the same interaction reproduces the full Fermi-surface scattering amplitude, it also provides a controlled input for Eliashberg- and Boltzmann-type calculations, in which pairing tendencies, quasiparticle lifetimes, and electron-electron-limited thermal transport depend directly on the angular structure of the scattering amplitude. More generally, the structure of sKO$^+$ may help guide the construction of low-energy effective field theories for correlated electron systems.

Looking forward, our VDMC data offer crucial numerical benchmarks for emerging experimental techniques that probe two-particle correlations in materials~\cite{PPS,Su_proposeARPES,fresch2023two}, and pave the way for next-generation \emph{ab initio} material-specific extensions, where downfolding the UEG vertex function may provide one route toward constructing accurate effective Hamiltonians. Extending this direction beyond the nearly-free-electron limit will require a material-specific treatment of band structure and core-electron effects, for example through effective-field-theory approaches to band renormalization~\cite{Cai2026CoreEFT}.

\emph{Acknowledgments.---} K.C. extends sincere thanks to
Boris Svistunov
and Carl A. Kukkonen for their valuable discussions.  K. C. was supported by the National Key Research and Development Program of China, Grant No. 2024YFA1408604, the National Natural Science Foundation of China under Grants No. 12474245 and
No. 12447103, and the GHfund A(202407010637). Z.L., P.H., and Y.D. were supported by the National Natural Science Foundation of China (under Grant No. 12275263), the Innovation Program for Quantum Science and Technology (under Grant No. 2021ZD0301900), and the Natural Science Foundation of Fujian Province of China (under Grant No. 2023J02032). 

\bibliography{ref}

\end{document}


\title{Supplemental Material}
\maketitle


\renewcommand{\thefigure}{S\arabic{figure}}
\renewcommand{\thetable}{S\arabic{table}}
\renewcommand{\theequation}{S\arabic{equation}}
\renewcommand{\thepage}{\arabic{page}}
\renewcommand{\thesection}{S\Roman{section}}

\setcounter{figure}{0}
\setcounter{table}{0}
\setcounter{equation}{0}
\setcounter{page}{1} 


\section{Variational Diagrammatic Monte Carlo Method and Convergence Analysis}
In this section, we provide the theoretical foundation and technical details of the VDMC framework employed in the main text.

Our investigation is grounded in the standard Hamiltonian for the uniform electron gas (UEG). In the second-quantization formalism, the system consists of interacting electrons and is described by the Hamiltonian:
\begin{equation}
    H = \sum_{\mathbf{k}\sigma} (\mathbf{k}^2 - \mu) \psi^\dagger_{\mathbf{k}\sigma} \psi_{\mathbf{k}\sigma} + \frac{1}{2} \sum_{\substack{\mathbf{q} \neq 0 \\ \mathbf{k}\mathbf{k}'\sigma\sigma'}} v_q \psi^\dagger_{\mathbf{k}+\mathbf{q}\sigma} \psi^\dagger_{\mathbf{k}'-\mathbf{q}\sigma'} \psi_{\mathbf{k}'\sigma'} \psi_{\mathbf{k}\sigma},
    \label{eq:UEG_Hamiltonian}
\end{equation}
where $\psi_{\mathbf{k}\sigma}$ and $\psi^\dagger_{\mathbf{k}\sigma}$ are the annihilation and creation operators for an electron with momentum $\mathbf{k}$ and spin $\sigma$, and $\mu$ denotes the chemical potential determined by the density parameter $r_s$. A significant challenge in addressing the many-body problem of this Hamiltonian arises from the long-range nature of the bare Coulomb interaction $v_q$. Standard diagrammatic expansions suffer from infrared divergences due to this long-range interaction~\cite{VanHoucke2020}. VDMC avoids this difficulty by expanding the theory with a renormalized chemical potential $\mu_R$ and a statically screened Yukawa interaction~\cite{chen2019combined,haule2022single,hou2024feynman,PhysRevB.93.161102,peskin2018introduction}, $v_q(\lambda_R)=4\pi e^2/(q^2+\lambda_R)$, while treating the difference from the physical Coulomb problem as counterterms. In this reorganized expansion, the reference propagator is $G_0(\mu_R)$ and the interaction line is $v_q(\lambda_R)$; the original Coulomb theory is recovered after the bookkeeping parameter introduced below is set to $\xi=1$.

The two counterterms have distinct physical roles. The chemical-potential counterterm $\delta\mu\equiv\mu-\mu_R$ is expanded order by order in the renormalized-order parameter $\xi$, $\delta\mu(\xi)=\sum_{p\ge 1}\xi^p\delta\mu^{(p)}$, and its coefficients are fixed by requiring that the interacting Fermi surface remain unshifted at each order. The interaction counterterm is obtained by reexpanding the bare Coulomb interaction as $4\pi e^2/q^2=v_q(\lambda_R)\sum_{n=0}^{\infty}[\xi\,\lambda_R v_q(\lambda_R)/(4\pi e^2)]^n$. Thus $\lambda_R$ acts as a variational screening parameter rather than a physical approximation: finite-order results are evaluated in the plateau region of $\lambda_R$, and the physical Coulomb interaction is restored at $\xi=1$.

After this reorganization, each renormalized order contains skeleton diagrams built from $G_0(\mu_R)$ and $v_q(\lambda_R)$ together with all chemical-potential and interaction-counterterm insertions. We represent these Feynman diagrams as computational graphs. Diagram generation and numerical-kernel construction are carried out within our FeynmanDiagram.jl framework~\cite{hou2024feynman}, which translates renormalized diagram classes into optimized representations compatible with Taylor-mode automatic differentiation (AD). Taylor-mode AD then generates the counterterm series coefficients directly on the diagrammatic computational graph, avoiding a separate manual enumeration of counterterm diagrams. The resulting high-dimensional integrals are sampled stochastically with the MCIntegration.jl framework~\cite{MCIntegrationjl}, and GPU acceleration can be implemented within this framework. This automated renormalized expansion is the basis for the sixth-order self-energy and four-point-vertex calculations used in the main text.

\begin{figure}
    \centering
    \nolinenumbers
    \includegraphics[width=0.4\linewidth]{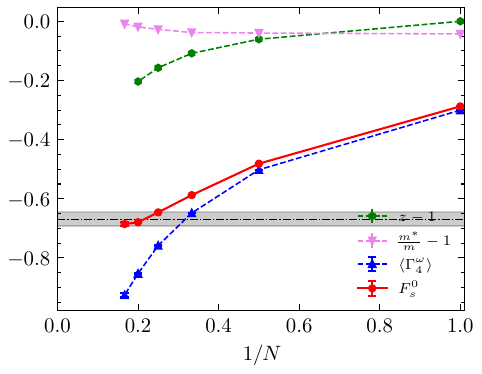}
    \caption{Convergence of the Landau parameter $F_s^0$ and its constituent parts---the quasiparticle weight ($z-1$), the effective mass ($(m^*/m)-1$), and the four-point vertex ($\langle\Gamma_4^\omega\rangle$)---as a function of inverse perturbation order $1/N$ for the UEG at $r_s=4.0$ with $\lambda_R=0.625$. The combined quantity $F_s^0$ converges significantly faster than its individual components.}
    \label{fig:convergence_of_landau_parameter}
\end{figure}
 
In this work, we developed an efficient technique to achieve rapid convergence in our calculations. We illustrate this method using the Landau parameter $F_0^s$, a key quantity discussed in the main text. This parameter relates to the quasiparticle weight $z$, the effective mass ratio $m^*/m$, and the angular average of the four-point vertex function $\langle\Gamma_4^\omega\rangle$. As defined in the main text, the vertex function takes the form $\Gamma_4^\omega = \lim_{\mathbf{q}\to 0} [\Gamma_4^{\sigma\sigma'}(\mathbf{k}_1\omega_{-1},\mathbf{k}_2\omega_0;\mathbf{k}_1+\mathbf{q}\omega_{0}) - v_q]$. The Landau parameter is then given by the relation $F_0^s = z^2(m^*/m)\langle\Gamma_4^\omega\rangle$, where $\langle\dots\rangle$ denotes the average over the angle between $\mathbf{k}_1$ and $\mathbf{k}_2$ on the Fermi surface. We compute the perturbative expansions of these constituent quantities in powers of a formal parameter $\xi$ using the same variational screening parameter $\lambda_R$:
\begin{align}
    z &= 1 + z_1\xi + z_2\xi^2 + \dots, \notag \\
    m^*/m &= 1 + m_1\xi + m_2\xi^2 + \dots, \\
    \langle\Gamma_4^\omega\rangle &= \gamma_1\xi + \gamma_2\xi^2 + \dots, \notag
\end{align}
where $z$ is derived from the numerical calculation of the self-energy diagram and $m^*/m$ is derived from the calculation of $F_1^s$. Specifically, $m^*/m = 1/(1-z^2\langle\Gamma^\omega_4\cos\theta\rangle)$.
Substituting these into the expression for the Landau parameter yields a series expansion $F_0^s = F^{(1)}\xi + F^{(2)}\xi^2 + \dots$, where the coefficients are determined order by order. Specifically, the first and second-order coefficients are given by:
\begin{align}
    F^{(1)} &= \gamma_1, \\
    F^{(2)} &= \gamma_2 + (2z_1\gamma_1 + m_1\gamma_1).
\end{align}
The rapid convergence observed in the final Landau parameter is a consequence of cancellations between same-order expansion terms generated by multiplying the constituent quantities in $F_0^s=z^2(m^*/m)\langle\Gamma_4^\omega\rangle$. For instance, in the second-order coefficient $F^{(2)}$, the direct vertex correction $\gamma_2$ is typically negative, whereas the terms generated by the quasiparticle-weight and effective-mass factors, $2z_1\gamma_1 + m_1\gamma_1$, are typically positive. These contributions partially cancel at the same perturbative order, reducing the magnitude of $F^{(2)}$ and accelerating the convergence of the final observable, as shown in Fig.~\ref{fig:convergence_of_landau_parameter}. Thus the cancellation is not imposed diagram by diagram; it emerges only after the self-energy, effective-mass, and vertex ingredients are combined into the physical Landau parameter. This explains why the individual ingredients may converge more slowly than the final observable quoted in the main text.

\section{Numerical results of the Landau parameters}
In this section, we present the precise numerical values of the Landau parameters derived from our VDMC calculations, which are summarized in Table~\ref{LPtable}.

The reduced temperature used in the main text is $\Theta=T/T_F$, with
\begin{equation}
T_F=\frac{\hbar^2 k_F^2}{2m_ek_B}, \qquad
k_F=\frac{(9\pi/4)^{1/3}}{a_0r_s}.
\end{equation}
Equivalently, $T_F\simeq 5.82\times10^5/r_s^2$ K for the electron gas. The simulated window $r_s=1$--6 and $\Theta=0.025$--0.05 is therefore a low-temperature, degenerate Fermi-liquid regime, with the lowest simulated temperatures reaching the order of room temperature at the largest $r_s$ values. The comparison between $\Theta=0.025$ and $0.05$ in the scattering-amplitude figure of the main text confirms that the forward-scattering amplitudes are essentially in the low-temperature limit, while the Cooper channel retains the expected logarithmic temperature dependence.

As discussed in the main text, the symmetric Landau parameter $F_0^s$ plays a pivotal role in characterizing the density-driven crossover from underscreening to overscreening. However, compared to the higher angular momentum components and the antisymmetric channels, $F_0^s$ exhibits relatively slower convergence, particularly in the strong coupling regime. To ensure the numerical stability and robustness of this critical quantity, we extended our diagrammatic expansion to the sixth order for the calculation of $F_0^s$. For the other Landau parameters ($F_{l>0}^s$ and $F_l^a$), a truncation at the fifth order was found to be sufficient for our purposes. 

\begin{table*}[h]
\nolinenumbers
\begin{tabular}{l|llllll}
\hline
$r_s$   & ~1.0           & ~2.0          & ~3.0           & ~4.0           & ~5.0         & ~6.0         \\ \hline
$F_0^s$ & -0.210(2)   & -0.367(6)  & -0.51(2)    & -0.67(2)    & -0.84(4)  & -1.06(8)  \\
$F_1^s$ & -0.049(3)   & -0.047(4)  & -0.040(3)   & -0.030(4)   & -0.024(3) & -0.015(4) \\
$F_2^s$ & -0.028(1)   & -0.030(3)  & -0.030(2)   & -0.029(2)   & -0.030(4) & -0.031(5) \\ \hline
$F_0^a$ & -0.175(1)   & -0.261(5)  & -0.320(9)   & -0.362(5)   & -0.400(5) & -0.43(2)  \\
$F_1^a$ & -0.066~4(4) & -0.088(5)  & -0.103(3)   & -0.115(7)   & -0.124(5) & -0.135(6) \\
$F_2^a$ & -0.019~6(6) & -0.012~4(6) & -0.006~0(4) & -0.001~3(4) & \ 0.001(4)  & \ 0.002(4)  \\ \hline
\end{tabular}
\caption{Numerical results of the Landau parameters.}
\label{LPtable}
\end{table*}
 
\section{Microscopic insights for the relation between the Fermi liquid theory and linear response theory}
In this section, we provide a detailed microscopic derivation of the density-density correlation function and the effective electron-electron interaction. By bridging the diagrammatic definition of the response function with microscopic theory, we clarify the physical origin of the relation between the Landau parameter and the exchange-correlation kernel, Eq.~(3) in the main text, discovered by our numerical results.

\begin{figure}[h]
\nolinenumbers
    \centering
    \tikzset{every picture/.style={line width=0.75pt}} 

\begin{tikzpicture}[x=0.75pt,y=0.75pt,yscale=-1,xscale=1]
\nolinenumbers

\draw [line width=0.75]    (243.33,139.62) .. controls (249.27,135.77) and (255.09,132.93) .. (260.9,131.42) .. controls (264.05,130.59) and (267.2,130.16) .. (270.37,130.16) .. controls (278.65,130.16) and (287.08,133.12) .. (295.99,140.05)(244.96,142.13) .. controls (250.61,138.48) and (256.13,135.76) .. (261.66,134.32) .. controls (264.56,133.56) and (267.45,133.16) .. (270.37,133.16) .. controls (278.07,133.16) and (285.88,135.98) .. (294.15,142.42) ;
\draw [shift={(274.77,131.95)}, rotate = 182.64] [fill={rgb, 255:red, 0; green, 0; blue, 0 }  ][line width=0.08]  [draw opacity=0] (8.93,-4.29) -- (0,0) -- (8.93,4.29) -- cycle    ;
\draw [line width=0.75]    (245.09,139.71) .. controls (252.82,145.99) and (261.99,149.47) .. (271.18,149.68) .. controls (271.47,149.69) and (271.76,149.69) .. (272.05,149.69) .. controls (280.9,149.69) and (289.7,146.62) .. (297.16,140.1)(243.2,142.04) .. controls (251.47,148.76) and (261.29,152.46) .. (271.11,152.68) .. controls (271.42,152.69) and (271.73,152.69) .. (272.05,152.69) .. controls (281.59,152.69) and (291.09,149.4) .. (299.13,142.36) ;
\draw [shift={(264.89,150.55)}, rotate = 5.92] [fill={rgb, 255:red, 0; green, 0; blue, 0 }  ][line width=0.08]  [draw opacity=0] (8.93,-4.29) -- (0,0) -- (8.93,4.29) -- cycle    ;
\draw    (120.03,140.97) .. controls (130.14,125.59) and (132.14,150.59) .. (140.14,139.59) .. controls (148.14,128.59) and (151.14,153.59) .. (161.14,139.59) ;
\draw    (301.21,141.23) .. controls (311.33,125.85) and (313.33,150.85) .. (321.33,139.85) .. controls (329.33,128.85) and (334.21,153.16) .. (344.21,139.16) ;
\draw  [fill={rgb, 255:red, 0; green, 0; blue, 0 }  ,fill opacity=1 ] (295.07,141.23) .. controls (295.07,139.54) and (296.45,138.16) .. (298.14,138.16) .. controls (299.84,138.16) and (301.21,139.54) .. (301.21,141.23) .. controls (301.21,142.93) and (299.84,144.3) .. (298.14,144.3) .. controls (296.45,144.3) and (295.07,142.93) .. (295.07,141.23) -- cycle ;
\draw    (19.03,142.97) .. controls (29.14,127.59) and (31.14,152.59) .. (39.14,141.59) .. controls (47.14,130.59) and (50.14,155.59) .. (60.14,141.59) ;
\draw  [fill={rgb, 255:red, 155; green, 155; blue, 155 }  ,fill opacity=0.87 ] (60.14,140.97) .. controls (60.14,132.41) and (73.55,125.47) .. (90.08,125.47) .. controls (106.62,125.47) and (120.03,132.41) .. (120.03,140.97) .. controls (120.03,149.53) and (106.62,156.47) .. (90.08,156.47) .. controls (73.55,156.47) and (60.14,149.53) .. (60.14,140.97) -- cycle ;
\draw    (201.14,142.95) .. controls (211.26,127.56) and (213.26,152.56) .. (221.26,141.56) .. controls (229.26,130.56) and (234.14,154.87) .. (244.14,140.87) ;
\draw   (460.03,134.97) .. controls (460.03,123.93) and (468.98,114.97) .. (480.03,114.97) .. controls (491.07,114.97) and (500.03,123.93) .. (500.03,134.97) .. controls (500.03,146.02) and (491.07,154.97) .. (480.03,154.97) .. controls (468.98,154.97) and (460.03,146.02) .. (460.03,134.97) -- cycle ;
\draw [line width=0.75]    (497.33,126.57) .. controls (503.25,122.74) and (509.05,120.88) .. (514.86,120.63) .. controls (515.31,120.61) and (515.77,120.6) .. (516.22,120.6) .. controls (527.18,120.6) and (538.21,126.3) .. (550.06,135.52)(498.96,129.09) .. controls (504.37,125.59) and (509.68,123.86) .. (514.99,123.63) .. controls (515.4,123.61) and (515.81,123.6) .. (516.22,123.6) .. controls (526.62,123.6) and (537.01,129.17) .. (548.22,137.89) ;
\draw [shift={(529.83,124.95)}, rotate = 197.91] [fill={rgb, 255:red, 0; green, 0; blue, 0 }  ][line width=0.08]  [draw opacity=0] (8.93,-4.29) -- (0,0) -- (8.93,4.29) -- cycle    ;
\draw [line width=0.75]    (499.09,140.67) .. controls (505.45,145.83) and (512.3,148.19) .. (519.25,148.19) .. controls (520.65,148.19) and (522.05,148.09) .. (523.45,147.91) .. controls (531.93,146.78) and (540.4,142.36) .. (548.16,135.58)(497.2,142.99) .. controls (504.16,148.65) and (511.66,151.19) .. (519.25,151.19) .. controls (520.78,151.19) and (522.31,151.08) .. (523.84,150.88) .. controls (532.85,149.69) and (541.88,145.05) .. (550.13,137.83) ;
\draw [shift={(518.14,149.67)}, rotate = 356.93] [fill={rgb, 255:red, 0; green, 0; blue, 0 }  ][line width=0.08]  [draw opacity=0] (8.93,-4.29) -- (0,0) -- (8.93,4.29) -- cycle    ;
\draw [line width=0.75]    (408.33,132.03) .. controls (414.09,128.3) and (419.75,125.09) .. (425.4,122.98) .. controls (429.91,121.31) and (434.42,120.33) .. (438.98,120.33) .. controls (446.13,120.33) and (453.43,122.71) .. (461.06,128.65)(409.96,134.55) .. controls (415.53,130.94) and (420.98,127.83) .. (426.44,125.8) .. controls (430.61,124.25) and (434.77,123.33) .. (438.98,123.33) .. controls (445.55,123.33) and (452.22,125.57) .. (459.22,131.01) ;
\draw [shift={(439.31,121.83)}, rotate = 175.34] [fill={rgb, 255:red, 0; green, 0; blue, 0 }  ][line width=0.08]  [draw opacity=0] (8.93,-4.29) -- (0,0) -- (8.93,4.29) -- cycle    ;
\draw [line width=0.75]    (410.09,132.12) .. controls (417.91,138.48) and (426.46,144.02) .. (435.02,146.23) .. controls (437.53,146.88) and (440.04,147.24) .. (442.53,147.24) .. controls (448.26,147.24) and (453.88,145.32) .. (459.16,140.7)(408.2,134.45) .. controls (416.37,141.09) and (425.33,146.83) .. (434.27,149.14) .. controls (437.03,149.85) and (439.79,150.24) .. (442.53,150.24) .. controls (448.93,150.24) and (455.23,148.12) .. (461.13,142.96) ;
\draw [shift={(427.57,145.16)}, rotate = 20.52] [fill={rgb, 255:red, 0; green, 0; blue, 0 }  ][line width=0.08]  [draw opacity=0] (8.93,-4.29) -- (0,0) -- (8.93,4.29) -- cycle    ;
\draw  [fill={rgb, 255:red, 0; green, 0; blue, 0 }  ,fill opacity=1 ] (241.07,142.23) .. controls (241.07,140.54) and (242.45,139.16) .. (244.14,139.16) .. controls (245.84,139.16) and (247.21,140.54) .. (247.21,142.23) .. controls (247.21,143.93) and (245.84,145.3) .. (244.14,145.3) .. controls (242.45,145.3) and (241.07,143.93) .. (241.07,142.23) -- cycle ;
\draw  [fill={rgb, 255:red, 0; green, 0; blue, 0 }  ,fill opacity=1 ] (406.07,135.23) .. controls (406.07,133.54) and (407.45,132.16) .. (409.14,132.16) .. controls (410.84,132.16) and (412.21,133.54) .. (412.21,135.23) .. controls (412.21,136.93) and (410.84,138.3) .. (409.14,138.3) .. controls (407.45,138.3) and (406.07,136.93) .. (406.07,135.23) -- cycle ;
\draw    (363.07,137.3) .. controls (373.19,121.92) and (375.19,146.92) .. (383.19,135.92) .. controls (391.19,124.92) and (396.07,149.23) .. (406.07,135.23) ;
\draw    (549.14,136.71) .. controls (559.26,121.32) and (561.26,146.32) .. (569.26,135.32) .. controls (577.26,124.32) and (582.14,148.63) .. (592.14,134.63) ;

\draw (167.14,130.59) node [anchor=north west][inner sep=0.75pt]  [font=\normalsize] [align=left] {$\displaystyle =$};
\draw (265.14,97.97) node [anchor=north west][inner sep=0.75pt]  [font=\footnotesize] [align=left] {$ $};
\draw (268.14,165.97) node [anchor=north west][inner sep=0.75pt]  [font=\footnotesize] [align=left] {$\displaystyle P$};
\draw (256.14,98.97) node [anchor=north west][inner sep=0.75pt]  [font=\footnotesize] [align=left] {$\displaystyle P+Q$};
\draw (133,116) node [anchor=north west][inner sep=0.75pt]  [font=\footnotesize] [align=left] {$\displaystyle Q$};
\draw (81,135) node [anchor=north west][inner sep=0.75pt]   [align=left] {$\displaystyle \chi $};
\draw (468.14,130.97) node [anchor=north west][inner sep=0.75pt]   [align=left] {$\displaystyle \Gamma _{4}$};
\draw (346.14,127.59) node [anchor=north west][inner sep=0.75pt]  [font=\normalsize] [align=left] {$\displaystyle +$};
\draw (513.14,89.97) node [anchor=north west][inner sep=0.75pt]  [font=\footnotesize] [align=left] {$ $};
\draw (516.14,157.97) node [anchor=north west][inner sep=0.75pt]  [font=\footnotesize] [align=left] {$\displaystyle P^{\prime }$};
\draw (501.14,96.97) node [anchor=north west][inner sep=0.75pt]  [font=\footnotesize] [align=left] {$\displaystyle P^{\prime } +Q$};
\draw (400.14,97.97) node [anchor=north west][inner sep=0.75pt]  [font=\footnotesize] [align=left] {$ $};
\draw (418.14,156.97) node [anchor=north west][inner sep=0.75pt]  [font=\footnotesize] [align=left] {$\displaystyle P-Q$};
\draw (433.14,98.97) node [anchor=north west][inner sep=0.75pt]  [font=\footnotesize] [align=left] {$\displaystyle P$};

\end{tikzpicture}
    \caption{Feynman diagrams for the Eliashberg equation.}
    \label{fig:PiDe}
\end{figure}
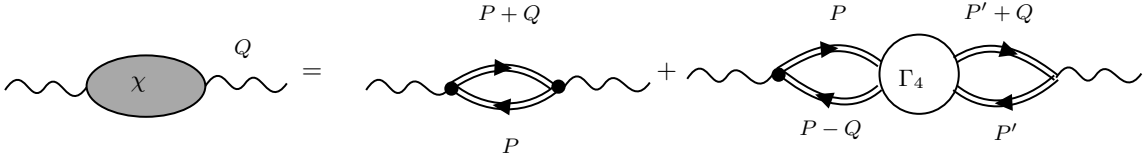

We begin with the density-density correlation function $\chi(q)$ in the imaginary-time formalism:
\begin{equation}
\chi(q) = \int_0^\beta\left<\mathcal{T}\hat{\rho}(\bq, \tau)\hat{\rho}(\bq, 0)\right> e^{i\omega_n \tau} d\tau,
\label{definitionofPi}
\end{equation}
where $\mathcal{T}$ is the time-ordering operator, the density operator $\hat{\rho}(\bq,\tau) = \sum_{\bp,\sigma}\hat{c}^\dagger_{\bp+\bq,\sigma}(\tau)\hat{c}_{\bp,\sigma}(\tau)$ and $q \equiv (i\omega_n,\bq)$ is the Matsubara-frequency/momentum vector. Through the Eliashberg equation~\cite{eliashberg1962transport} as shown in Fig.~\ref{fig:PiDe}, $\chi(q)$ is related to the four-point vertex function $\Gamma^4(p,p^\prime,k)$ and the quasiparticle Green's function $G(p)$ by the equation

\begin{equation}
    \chi(q) = \sum_{p\sigma} G_\sigma(p+q)G_\sigma(p) + \sum_{p\sigma,p^\prime\sigma^\prime}G_\sigma(p+q)G_\sigma(p)\Gamma^4_{\sigma\sigma^\prime}(p,p^\prime,q)G_{\sigma^\prime}(p^\prime)G_{\sigma^\prime}(p^\prime-q),
\end{equation}
Near the Fermi surface, the Green's function takes the form $G_\sigma(p) \approx \frac{z}{i\epsilon_n - v_F(|\mathbf{p}|-k_F)}$, where $v_F$ is the Fermi velocity and $z$ is the quasiparticle renormalization factor.

For small momentum and energy transfer ($q \to 0$), the product of Green's functions $g_\sigma(p,q) \equiv G_\sigma(p)G_\sigma(p+q)$ exhibits a singularity arising from the Fermi surface poles. It can be decomposed into a singular part $\Pi$ and a regular part $\phi$:
\begin{equation}
    g_\sigma(p,q)\equiv G_\sigma(p)G_\sigma(p+q) \sim \Pi(p,q)\delta(i\epsilon_n)\delta(|\bp|-k_F)+\phi(p),
    \label{gstruct}
\end{equation}
The first term captures the particle-hole excitations restricted to the Fermi surface, while $\phi(p)$ represents contributions from high-energy excitations. To simplify the derivation, we adopt a matrix notation where multiplication implies summation over momentum $p$ and spin $\sigma$. The correlation function can be compactly written as:
\begin{equation}
    \chi(q) = \mathrm{Tr}[g(1+\Gamma_4 g)].
\end{equation}

We introduce the quasiparticle irreducible vertex function $\Gamma^\omega$, defined in the limit where $q \to 0$ with $|\mathbf{q}|/\omega_n \to 0$. This quantity is related to the full vertex $\Gamma_4$ through a Dyson-like equation involving the singular part $\Pi$:
\begin{equation}
    \Gamma_4 = \Gamma^\omega + \Gamma^\omega \Pi \Gamma_4.
    \label{Gamma4o}
\end{equation}
Crucially, in Landau Fermi liquid theory~\cite{landau1959theory}, this irreducible vertex $\Gamma^\omega$ is directly related to the Landau quasiparticle interaction $f_{\sigma\sigma^\prime}(\theta)$ on the Fermi surface:
\begin{equation}
    f_{\sigma\sigma^\prime}(\theta) = z^2 \Gamma^\omega_{\sigma\sigma^\prime}(\mathbf{p}_F, \mathbf{p}^\prime_F).
\end{equation}

To solve for the response function, we employ a matrix identity introduced in Ref.~\cite{PhysRev.140.A1869}: for matrices satisfying $f = f_1 + f_2$ and $S = S_0 + S_0 f_2 S$, one can derive:
\begin{equation}
    f(1+Sf) = f_1(1+S_0 f_1) + (1+f_1 S_0) f_2 (1-S_0 f_2)^{-1} (1+S_0 f_1).
\end{equation}
Identifying $f \to g$, $S \to \Gamma_4$, and using the decomposition $g = \phi + \Pi$, we utilize the Ward identity relation $1 + \sum \phi \Gamma^\omega = 1/z$~\cite{abrikosov2012methods} to simplify the resulting expression. This leads to:
\begin{equation}
    \chi(q) = \Phi + \frac{1}{z^2}\mathrm{Tr}[\Pi(1-\Gamma^\omega \Pi)^{-1}],
\end{equation}
where $\Phi = \mathrm{Tr}[\phi(1+\phi\Gamma^\omega)]$. It can be further shown that the regular contribution $\Phi$ vanishes ($\Phi=0$). Thus, the response function is determined solely by the Fermi surface contributions:
\begin{equation}
    \chi(q) = \frac{1}{z^2} \sum_{\theta\sigma} \frac{\Pi(\theta)}{1 - \sum_{\theta^\prime\sigma^\prime} \Gamma^\omega_{\sigma\sigma^\prime}(\theta, \theta^\prime)\Pi(\theta^\prime)},
\end{equation}
where $\theta$ and $\theta^\prime$ denote the angles between the transfer momentum $\mathbf{q}$ and the momenta of the incoming quasiparticles $\mathbf{p}$ and $\mathbf{p}^\prime$, respectively.

Substituting the Landau parameter relation $\Gamma^\omega = z^{-2} f$, the equation becomes:
\begin{equation}
    \chi(q) = \frac{1}{z^2} \sum_{\theta,\sigma} \frac{\Pi(\theta)}{1 - \sum_{\theta^\prime\sigma^\prime} z^{-2} f_{\sigma\sigma^\prime}(\theta, \theta^\prime)\Pi(\theta^\prime)}.
\end{equation}
In the static limit $q=(0, \mathbf{q})$, the term $\sum \Pi(\theta)$ corresponds to the renormalized Lindhard function: $\sum \Pi = z^2 (m^*/m) \Pi_0(q)$.

Supported by the numerical results shown in Figs.~2 and~3 in the main text, the angle dependence of the Landau quasiparticle interaction in the UEG is sufficiently weak that the interaction is dominated by the zeroth-order term ($f_0^s \gg f_l^s$ for $l\geq1$). We can therefore approximate it using the zeroth-order Landau parameter $f_0^s = F_0^s/N_F^*$, where $N_F^*$ denotes the quasiparticle density of states. By explicitly including the long-range Coulomb interaction $v_q = 4\pi e^2/q^2$---which enters the singlet channel as $v_q + f_0^s$ in Landau-Silin theory~\cite{silin1958theory, silin1958theory2}---the response function simplifies to:
\begin{equation}
\chi(q) \approx \frac{\frac{m^*}{m}\Pi_0(q)}{1 - \frac{m^*}{m} (v_q + f_0^s) \Pi_0(q)}.
\end{equation}
This expression takes a form similar to the response function defined by the exchange-correlation (XC) kernel $f_{\text{XC}}$, written as $\chi = \Pi_0(q) / [1 - (v_q + f_{\text{XC}}(q))\Pi_0(q)]$. Furthermore, our numerical results indicate that the effective mass ratio $m^*/m$ approaches unity in the metallic regime. Based on this, we validate the local density approximation (LDA) and the relationship presented in Eq.~(3) of the main text:
\begin{equation}
f_{\text{XC}}(q)\approx f_{\text{XC}}(q=0) \approx \frac{F_0^s}{N_F^*},
\end{equation}
for the metallic gas.

By revisiting Eq.~\eqref{Gamma4o}, we can derive that 
\begin{equation}
    \Gamma_4 = \Gamma_4^\omega(1-\Gamma_4^\omega\Pi)^{-1},
    \label{Gamma4equation}
\end{equation}
which encodes the singular part of the two-electron correlation structure as $\omega/q \to 0$. We can then obtain the effective electron-electron interaction within LDA and Landau-Silin theory as 
\begin{equation}
    R(q)\sim \frac{v_q+f_0^s}{1-\frac{m^*}{m}(v_q+f_0^s)\Pi_0(q)}.
\end{equation}

To properly describe the effective interaction in the finite-$q$ regime, one must eliminate the double-counting that arises from the iterative summation of Feynman diagrams as shown in Eq.~\eqref{Gamma4equation}. A common procedure is to subtract the term $f_0^s$, which ensures that the double-counting is exactly canceled and the effective interaction correctly vanishes in the $q \to \infty$ limit:
\begin{equation}
 R(q) = \frac{v_q+f_0^s}{1-\frac{m^*}{m}(v_q+f_0^s)\Pi_0(q)}-f_0^s.   
\end{equation}
In the UEG limit where $m^* \approx m$, this result recovers the structure of the charge-based Kukkonen--Overhauser (KO) interaction~\cite{PhysRevB.20.550}. However, while this subtraction scheme is enforced by the asymptotic behavior at $q \to \infty$, it is not necessarily optimal for the physically relevant region below the Fermi energy. One of the central findings of our work is that for these sub-Fermi scales, the double-counting should instead be addressed using the $\text{sKO}^+$ form presented in Eqs.~(1) and~(2) of the main text.

\section{Crossover of the Coulomb screening effect}
\begin{figure}[h]
    \centering
    \subfloat[]{\includegraphics[width = 0.43\linewidth]{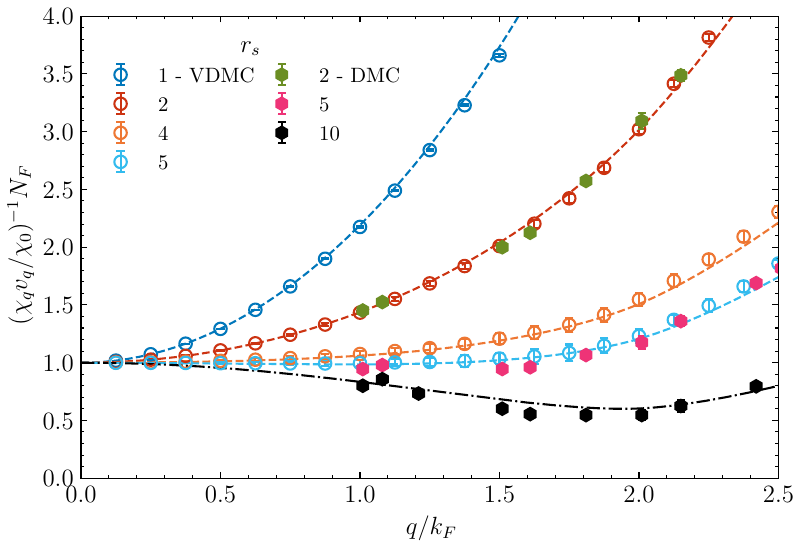} \label{fig:Density-correlation}}
    \subfloat[]{\includegraphics[width = 0.45\linewidth]{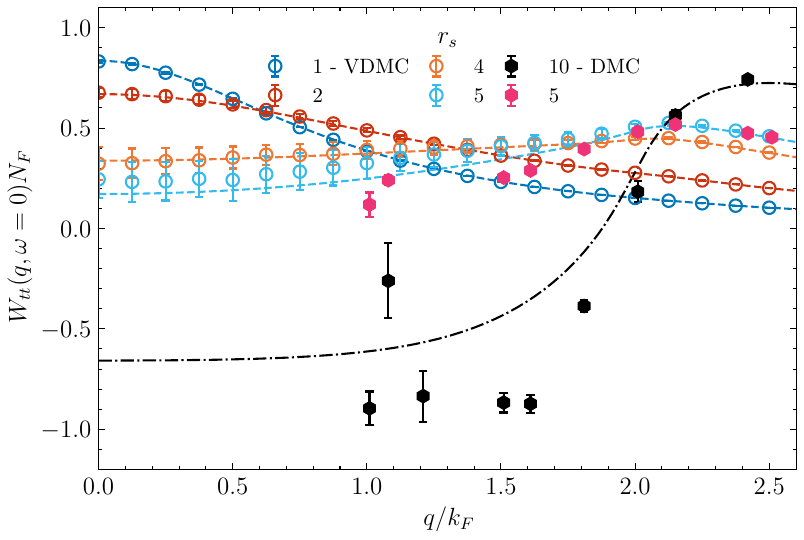}\label{fig:Wtt}}
    \caption{(a) The inverse normalized density-density response function, $(\chi_q v_q / \chi_0)^{-1} N_F$, calculated for various $r_s$. The symbols represent data from VDMC (open circles) and Diffusion Monte Carlo (DMC, filled symbols)~\cite{DMC3DResults}. The dashed lines correspond to the theoretical prediction using LDA. (b) The effective test charge-test charge interaction $W_{tt}(q, \omega=0)$ normalized by the Fermi density of states $N_F$. The transition from positive to negative values at small momenta signals the onset of the overscreening phase. The excellent agreement between the VDMC/DMC data and the LDA-based curves confirms the effectiveness of LDA.}
    \label{crossover}
\end{figure}
In this section, we provide further details on the density-driven crossover of the Coulomb screening effect discussed in the main text. To visualize this phenomenon, we employed VDMC to calculate the static density-density correlation function, $\chi(q)$, and the effective test charge-test charge interaction $W_{tt}(q)$, which relates to the ion-ion interaction in metals, as shown in Fig.~\ref{crossover}. In the linear response formalism, $W_{tt}(q)$ is expressed in terms of the bare Coulomb potential $v_q$, the non-interacting Lindhard polarizability $\Pi_0(q)$, and the static XC kernel $f_{\text{XC}}(q)$ as:
\[
W_{tt}(q) = \frac{v_q [1 - f_\text{XC}(q) \Pi_0(q)]}{1 - [v_q + f_\text{XC}(q)] \Pi_0(q)}.
\]

To disentangle many-body correlations from the Coulomb singularity, we analyzed the normalized response function $\chi(q)v_q/\chi_0(q)$. As shown in Fig.~\ref{fig:Density-correlation}, the microscopic VDMC results exhibit remarkable agreement with benchmark DMC data~\cite{DMC3DResults} and theoretical predictions using a static kernel $f_\text{XC}(q) = f_\text{XC}(0)$ (shown by the dashed lines). This concordance confirms that LDA effectively captures the essential screening physics in the small-$q$ regime.

The numerical results clearly manifest a physical crossover in the density regime $5 < r_s < 6$, consistent with the compressibility divergence ($F_0^s \to -1$). For metallic densities ($r_s < r_s^c$), the system resides in the conventional underscreening regime, characterized by positive static compressibility and a purely repulsive effective interaction. Conversely, as the density decreases ($r_s > r_s^c$), an overscreening phase emerges where the static compressibility and dielectric function become negative in the long-wavelength limit. Physically, this induces a counterintuitive attractive interaction between like-charged test particles at large distances, as evidenced by the negative tail in $W_{tt}(q)$ (Fig.~\ref{fig:Wtt}), while the interaction remains repulsive at short ranges. Additionally, Fig.~\ref{fig:Density-correlation} suggests that this crossover may be related to a stable RG fixed point, as $\chi_0N_F/\chi_qv_q$ approaches 1 as $q \to 0$ for any $r_s$.

\section{Results of the general polynomial expansion}
In this section, we detail the procedure for deriving the residual part of the scattering amplitude that extends beyond the charge-only KO$^+$ ansatz. To rigorously compare theoretical ansatzes with the numerical scattering amplitudes $A^{\sigma\sigma'}(\theta, \phi)$ computed via VDMC, we first establish the general mapping between effective interaction potentials $R(\mathbf{q})$ and the observable scattering amplitudes. This construction relies on the specific spin configuration and the requirements imposed by the Pauli exclusion principle.

For fermions with parallel spins ($\uparrow\uparrow$), the total wavefunction must be antisymmetric with respect to particle exchange. Consequently, the scattering amplitude is constructed as the difference between the direct interaction term, with momentum transfer $\mathbf{q} = \mathbf{k}_1 - \mathbf{k}_3$, and the exchange interaction term, with momentum transfer $\mathbf{q}' = \mathbf{k}_1 - \mathbf{k}_4$. The resulting amplitude takes the form:
\begin{equation}
    a^{\uparrow\uparrow}(\theta,\phi) = R^{\uparrow\uparrow}(\mathbf{k}_1-\mathbf{k}_3) - R^{\uparrow\uparrow}(\mathbf{k}_1-\mathbf{k}_4).
\end{equation}
In contrast, for the antiparallel spin channel ($\uparrow\downarrow$), the particles are distinguishable, and in the standard formulation of charge-channel models such as the KO$^+$ and our proposed sKO$^+$ ansatz, only the direct scattering process contributes to the amplitude:
\begin{equation}
    a^{\uparrow\downarrow}(\theta,\phi) = R^{\uparrow\downarrow}(\mathbf{k}_1-\mathbf{k}_3).
\end{equation}
It is worth noting that for the full KO$^\pm$ interaction, which explicitly incorporates a transverse spin-exchange interaction $R_{\text{KO}^-}$ mediated by spin fluctuations, the antiparallel amplitude acquires an additional contribution. In that specific case, the amplitude is modified to $a^{\uparrow\downarrow}_{\text{KO}^\pm}(\theta,\phi) = R^{\uparrow\downarrow}(\mathbf{k}_1-\mathbf{k}_3) - 2R_{\text{KO}^-}(\mathbf{k}_1-\mathbf{k}_4)$.

To systematically quantify the correlation terms missing from KO$^+$, we analyze the residual scattering amplitude, defined as $\delta A^{\sigma\sigma'}(\theta,\phi) = A^{\sigma\sigma'}_{\text{VDMC}}(\theta,\phi) - A^{\sigma\sigma'}_{\text{KO}^+}(\theta,\phi)$. We perform a general polynomial expansion of this residual part using a set of basis functions $X_{lk}(\theta,\phi)$ that are orthogonal on the Fermi surface~\cite{pfitzner1985general}:
\begin{equation}
    \delta A^{\sigma\sigma'}(\theta,\phi) = \sum_{k=0}^\infty \sum_{l=0}^k \delta A_{lk}^{\sigma\sigma'} X_{lk}(\theta,\phi).
\end{equation}
The basis functions are constructed from Legendre polynomials $P_l(x)$ and Jacobi polynomials $P_n^{(a,b)}(x)$ as $X_{lk}(\theta,\phi) = (-1)^l (k+1)^{\frac{1}{2}} (2l+1)^{\frac{1}{2}} \sin^{2l}(\theta/2) P_l(\cos\phi) P^{(2l+1,0)}_{k-l}(\cos \theta)$.

\begin{figure}[h]
    \centering
    \includegraphics[width=0.45\linewidth]{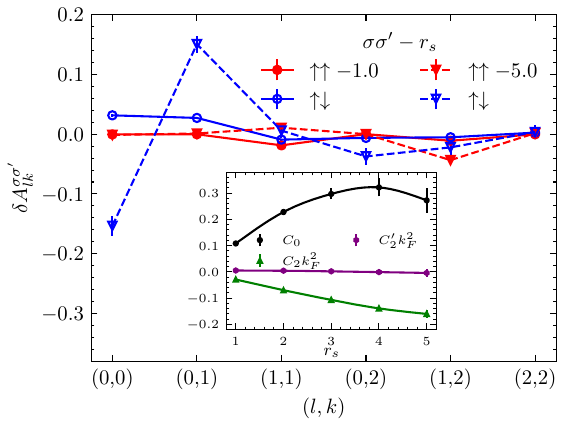}
    \caption{General polynomial expansion~\cite{pfitzner1985general} of the residual part of the scattering amplitude. Inset: Parametrization of the effective field action describing the electron-electron correlation beyond the KO$^+$ ansatz. Here, $C_0$ relates to the s-wave scattering length, $C_2$ relates to the s-wave effective range, and $C_2'$ relates to a negligible higher-order p-wave correction. The results show that the residual part can be described by an s-wave scattering process.}
    \label{fig:LDAKO_sp}
\end{figure}

As illustrated in Fig.~\ref{fig:LDAKO_sp}, the resulting expansion coefficients $\delta A_{lk}^{\sigma\sigma'}$ reveal a clear physical distinction between the spin channels. For parallel spins, the coefficients are negligible, confirming that the KO$^+$ ansatz provides a sufficient description of the interaction. However, for antiparallel spins, the residual is significantly non-zero and is clearly dominated by the isotropic $(l,k)=(0,0)$ and $(0,1)$ components. This observation justifies parameterizing the residual effective interaction, $\delta R^{\uparrow\downarrow}$, as a simple s-wave contact interaction, analogous to effective field theories for dilute Fermi systems~\cite{HAMMER2000277,WELLENHOFER2020135247}:
\begin{equation}
    \delta R^{\uparrow\downarrow}(\mathbf{k}_1,\mathbf{k}_2;\mathbf{k}_3,\mathbf{k}_4) = C_0 + C_2\frac{(\mathbf{k}_1-\mathbf{k}_2)^2+(\mathbf{k}_3-\mathbf{k}_4)^2}{2},
    \label{C0C2def}
\end{equation}
with the higher-order p-wave correction $C_2'(\mathbf{k}_1-\mathbf{k}_2)\cdot(\mathbf{k}_3-\mathbf{k}_4)$ negligible, as shown in Fig.~\ref{fig:LDAKO_sp}.
In this expression, the leading-order constant $C_0$, which is proportional to the s-wave scattering length, is determined by the dominant $\delta A_{00}^{\uparrow\downarrow}$ component. The next-to-leading-order constant $C_2$, related to the s-wave effective range, is fixed by the sub-dominant term $\delta A_{01}^{\uparrow\downarrow}$. This derived correction term $\delta R$ with its relevant parameters shown in the inset of Fig.~\ref{fig:LDAKO_sp} constitutes the basis of the sKO$^+$ ansatz employed in the main text. The fitted values of $C_0$ and $C_2$ for the metallic density range $r_s=1$--$5$ are summarized in Table~\ref{C0C2table}.

\begin{table*}[h]
\nolinenumbers
\begin{tabular}{l|r@{.}lr@{.}lr@{.}lr@{.}lr@{.}l}
\hline
$r_s$ & \multicolumn{2}{c}{1.0} & \multicolumn{2}{c}{2.0} & \multicolumn{2}{c}{3.0} & \multicolumn{2}{c}{4.0} & \multicolumn{2}{c}{5.0} \\ \hline
$C_0$ & 0&109(3) & 0&228(5) & 0&297(10) & 0&32(2) & 0&27(2) \\
$C_2$ & -0&029(1) & -0&070(2) & -0&107(3) & -0&139(5) & -0&160(7) \\ \hline
\end{tabular}
\caption{Fitted coefficients of the short-range s-wave correction $\delta R^{\uparrow\downarrow}$ in Eq.~\eqref{C0C2def} for $r_s=1$--$5$.}
\label{C0C2table}
\end{table*}

\section{Transport properties derived from the scattering amplitude}
As mentioned in the main text, the electron-electron scattering amplitude $a_{\sigma\sigma'}(\theta,\phi)$ allows us to derive the electron-electron contribution to several transport properties, including the electron-electron scattering rate $1/\tau$ and the thermal resistivity $W_{ee}$. Specifically, they are derived from the following formulas~\cite{mahan2013many,PhysRev.185.323,PhysRevLett.21.279}:
\begin{align}
    W_{ee} &= \frac{3(m^*)^4T}{8\pi^2\hbar^6k_F^3H(\lambda)}\langle \frac{\omega(\theta,\phi)(1-\cos\theta)}{\cos\frac{\theta}{2}}\rangle,\\
        B_{ee} &= \frac{3(m^*)^4}{8\pi^2\hbar^6k_F^3H(\lambda)}\langle \frac{\omega(\theta,\phi)(1-\cos\theta)}{\cos\frac{\theta}{2}}\rangle,\\
    \frac{1}{\tau} & = \frac{(m^*)^3}{16\pi^2\hbar^6}(k_BT)^2\langle \frac{\omega(\theta,\phi)}{\cos \frac{\theta}{2}}\rangle,
\end{align}
where 
\begin{align}
     &H(\lambda)= \frac{3-\lambda}{4}\sum_{n=0}^\infty \frac{4n+5}{(n+1)(2n+3)[(n+1)(2n+3)-\lambda]},\\
    &\lambda  = \frac{\langle\omega(\theta,\phi)(1+2\cos \theta)/\cos\frac{\theta}{2}\rangle}{\langle\omega(\theta,\phi)/\cos\frac{\theta}{2}\rangle}, \\
    &\omega(\theta,\phi) = \frac{2\pi}{\hbar}[\frac{1}{2}|a_{\uparrow\downarrow}(\theta,\phi)|^2+\frac{1}{4}|a_{\uparrow\uparrow}(\theta,\phi)|^2].
\end{align}
where $\langle\cdots\rangle = \frac{1}{4\pi}\int_0^{2\pi}d\phi\int_0^\pi d\theta \sin\theta$ denotes the average over the solid angle. 
For the comparison with elemental metals in the main text, we use the standard electron-gas mapping in which the valence-electron density fixes $r_s$ and the calculated UEG scattering amplitude is inserted into the Fermi-surface angular averages above. No realistic crystal structures or DFT band dispersions are included in this benchmark. This approximation is most controlled for nearly-free-electron metals with simple Fermi surfaces and small band-mass or core-polarization corrections. For materials with anisotropic Fermi surfaces, multiple bands, substantial $d$ or $f$ orbital character, or large dynamical mass renormalization, the same sKO$^+$ interaction should be used only after embedding it in a material-specific quasiparticle description.

As mentioned in the main text, the s-p approximation proposed by Dy and Pethick~\cite{transportcoeff1969,MacDonald_1980} approximates the electron-electron scattering amplitude using the s-wave and p-wave contributions and shows good agreement with experimental results. Specifically, the scattering amplitude can be written as
\begin{align}
    a^{\mathrm{sp}}_{\uparrow\uparrow}(\theta,\phi)
    &= \frac{\cos\phi}{N_F^*}
    \left[A_0^s+A_0^a+3(A_1^s+A_1^a)\cos\theta\right],\\
    a^{\mathrm{sp}}_{\uparrow\downarrow}(\theta,\phi)
    &= \frac{1}{2N_F^*}\Bigl\{
    \cos\phi\left[A_0^s+A_0^a+3(A_1^s+A_1^a)\cos\theta\right] \notag\\
    &\hspace{3.0cm}
    +A_0^s-3A_0^a+3(A_1^s-3A_1^a)\cos\theta
    \Bigr\}
\end{align}
where $A^{s,a}_l = \frac{F_l^{s,a}}{1+F_l^{s,a}}$ is derived from the Landau parameter.

\bibliographystyle{apsrev4-2}
\bibliography{ref}

\clearpage